\documentclass[12pt]{article}
\usepackage{graphicx}

\newwrite\ffile\global\newcount\figno \global\figno=1

\def\writedef#1{}

\input epsf
\def\figin{\epsfcheck\figin}\def\figins{\epsfcheck\figins}
\def\epsfcheck{\ifx\epsfbox\UnDeFiNeD
\message{(NO epsf.tex, FIGURES WILL BE IGNORED)}
\gdef\figin##1{\vskip2in}\gdef\figins##1{\hskip.5in}
\else\message{(FIGURES WILL BE INCLUDED)}%
\gdef\figin##1{##1}\gdef\figins##1{##1}\fi}

\def\figinsert{}
\def\ifig#1#2#3{\xdef#1{fig.~\the\figno}
\writedef{#1\leftbracket fig.\noexpand~\the\figno}%
\figinsert\figin{\centerline{#3}}\medskip\centerline{\vbox{\baselineskip12pt
\advance\hsize by -1truein\center\footnotesize{  Fig.~\the\figno.} #2}}
\bigskip\endinsert\global\advance\figno by1}
\def\endinsert{}

\usepackage{amssymb}
\usepackage{graphics}

\begin{document}
\baselineskip 18pt
\newcommand{\Tr}{\mbox{Tr\,}}
\newcommand{\beq}{\begin{equation}}
\newcommand{\eeq}{\end{equation}}
\newcommand{\bea}{\begin{eqnarray}}
\newcommand{\eea}[1]{\label{#1}\end{eqnarray}}
\renewcommand{\Re}{\mbox{Re}\,}
\renewcommand{\Im}{\mbox{Im}\,}

\def\N{{\cal N}}


\thispagestyle{empty}
\renewcommand{\thefootnote}{\fnsymbol{footnote}}

{\hfill \parbox{4cm}{
        SHEP-04-11 \\
}}

\bigskip

\begin{center} \noindent \Large \bf
Chiral Dynamics From AdS Space
\end{center}

\bigskip\bigskip\bigskip

\centerline{ \normalsize \bf Nick Evans and Jonathan P. Shock
\footnote[1]{\noindent \tt
 evans@phys.soton.ac.uk, jps@phys.soton.ac.uk} }

\bigskip
\bigskip\bigskip

\centerline{ \it Department of Physics}
\centerline{ \it Southampton University}
\centerline{\it  Southampton, S017 1BJ }
\centerline{ \it United Kingdom}
\bigskip

\bigskip\bigskip

\renewcommand{\thefootnote}{\arabic{footnote}}

\centerline{\bf \small Abstract}
\medskip

{\small \noindent We study the low energy dynamics of pions in a
gravity dual of chiral symmetry breaking. The string theory
construction consists of a probe D7 brane in the Constable Myers
non-supersymmetric background, which has been shown to describe
chiral symmetry breaking in the pattern of QCD. We expand the D7
brane's Dirac Born Infeld action for fluctuations that correspond
to the Goldstone mode and show that they take the form of a
non-linear chiral lagrangian. We numerically compute the quark
condensate, pion decay constant and higher order Gasser Leutwyler
coefficients. We find their form is consistent with naive
dimensional analysis estimates. We also explore the gauging of the
quark's chiral symmetries and the vector meson spectrum.}

\newpage


\section{Introduction}

The dynamics and phenomenology of QCD are dominated by quarks. In
particular the vast majority of known hadronic states can be
identified as having constituent quarks and the low energy
dynamics is controlled by the chiral symmetry breaking quark
condensate. The discovery of the AdS/CFT Correspondence \cite{Mal,
Gub, Wit} has raised the hope of providing a weakly coupled
gravitational description of the strong coupling phase of QCD.
Understanding fundamental representation quarks in this setting
must therefore be a priority.

Recently a simple mechanism for including quarks in the AdS/CFT
Correspondence has been found by Karch and Katz \cite{KarchKatz}.
A probe D7 brane is added to the original D3 brane construction of
the AdS/CFT Correspondence. The new ``37" open strings generate
quarks in the field theory on the D3 brane world volume. The world
volume theory has ${\cal N}=2$ supersymmetry. Karch and Katz
identified the additional ``77" open strings on the D7 world
volume as playing a holographically dual role to the gauge
invariant quark operators of the gauge theory. Treating the D7 as
a probe corresponds to quenching in the gauge theory. The dynamics
of quarks and their bound states have been studied in a number of
supersymmetric gauge field backgrounds using these techniques
\cite{Weiner} - \cite{Strass}.

If we wish to study chiral symmetry breaking in the pattern of QCD
we must look at a non-supersymmetric gauge theory since
supersymmetry forbids such a quark condensate. A number of
gravitational duals of non-supersymmetric backgrounds exist
\cite{Witten} - \cite{CM} and \cite{Volkov}-\cite{Kuperstein}. The
simplest cases involve the deformation of the original AdS/CFT
Correspondence by the inclusion of relevant operators in the field
theory \cite{Girardello}, which corresponds to switching on bulk
supergravity fields. In these cases the UV of the gauge theory
returns to ${\cal N}=4$ super Yang Mills and the operator
identification between the two dual theories is cleanly
understood. Such theories do not have total decoupling of the
super-partners that are given mass since the theory is strongly
coupled in the UV - these extra states have masses of order the
strong interaction scale $\Lambda$. Simply breaking supersymmetry
is sufficient to allow a quark condensate though.
Non-supersymmetric deformations \cite{Volkov}-\cite{Kuperstein} of
the Klebanov Strassler \cite{KS} and Maldecena Nunez \cite{MN}
gravity duals also exist but here the operator matching, even in
the UV, is less clear.

The first study of chiral symmetry breaking in this formalism was
made in \cite{us}. The deformed AdS$_5\times S^5$ geometry of
Constable and Myers \cite{CM}, which corresponds to the addition
of an R-chargeless dimension 4 operator such as Tr$F^{\mu \nu}
F_{\mu \nu}$ to the ${\cal N}=4$ theory, was used. The existence
of a condensate and massless pions in the limit where the quark
mass went to zero were identified. In this paper we will make
further study of that case. The Constable Myers' geometry has a
singularity in the interior, the precise significance of which is
unclear; the singularity might correspond to the presence of some
expanded D-brane set up in the interior and signal the strong
interaction scale $\Lambda$. It turns out that the core of the
geometry is repulsive to the D7 brane probe when quarks are
included and this is what triggers the chiral symmetry breaking in
the model. Pleasingly this repulsion also makes sure that the D7
branes avoid the central singularity so for the purposes of this
study we can set aside study of the singularity.

Chiral symmetry breaking by the same mechanism has also been
studied in \cite{Mateos2}. They use a geometry around D4 branes
wrapped on a circle which describes a 3+1 dimensional Yang Mills
like theory in the IR. This geometry has no interior singularity
but the core is again repulsive to D7 brane probes triggering
chiral symmetry breaking. In this case though the UV of the theory
becomes strongly coupled and wishes to become an M5 brane
construction. The universality of the IR mechanism is encouraging
and provides support for further study in the Constable Myers
background.

Here we will return to that Constable Myers configuration
\cite{CM,us} and examine chiral symmetry breaking and its
consequences in more detail. First we refine the numerical
analysis of \cite{us} by solving the Euler Lagrange equation,
describing how the D7 brane lies in the geometry, starting with
the appropriate regular infra-red boundary conditions. This allows
us to identify the regular physical flows without tuning. We
stress the geometrical description of chiral symmetry breaking
provided by the set up where the repulsion of the interior
geometry forces the D7 brane to lie in a symmetry breaking
configuration. For small quark mass the vacuum energy of the
configuration is proportional to the quark mass as expected in the
chiral lagrangian formalism \cite{Coleman, Georgi} (which we will
review in the next section). It is therefore possible to extract
the quark condensate which we show matches with the value obtained
in \cite{us} by looking at the UV boundary conditions on the
flows.

Next we move on to study fluctuations of the D7 about the vacuum
configuration, that describe the Goldstone mode, or pion fields.
We show that the lagrangian terms match those expected in the
chiral lagrangian and then compute the couplings. In particular
the pion mass is shown numerically to have a linear dependence on
the square root of the quark mass for low quark mass. It is then
possible to compute the pion decay constant which we show has the
correct dependence on the number of colours $N$ and has a
numerical suppression factor relative to the strong coupling scale
consistent with that seen in QCD and naive arguments
\cite{Coleman, Georgi, Georgi2}. In the chiral limit we then look
at terms in the low energy lagrangian involving four pion fields
and estimate their size. We again find a match to naive estimates
\cite{Gasser, Georgi2}.

A natural next step is to include multiple D7 probes and study the
resulting theory with $N_f$ quark flavours via the non-abelian DBI
action \cite{nadbi}. In fact the quarks couple to the adjoint
scalar field in the ${\cal N}=2$ UV gauge theory via a
superpotential term $\tilde{Q} A Q$ and enhancing the number of
quark flavours does not therefore enhance the chiral flavour
group. The adjoint scalar is though massive on the scale $\Lambda$
and we might expect an accidental symmetry in the IR. This
phenomena has been discussed in the context of \cite{Mateos2}. The
additional fields in the non-abelian DBI action are also massless
but have interaction terms that are not Goldstone like - these are
just the remnants of the usual commutator interactions of the
scalar fields on a brane. As a test of the Goldstone like nature
of these fields we compute the pion decay constant via lagrangian
terms that are not present in the abelian case, neglecting the
commutator interactions. We find excellent numerical agreement
with our previous value showing that these fields are rather
Goldstone like.

Finally we study the gauge field on the world volume of the D7
which is dual to weakly gauging the vector U(1) baryon number
symmetry and also the vector meson spectrum. We have been unable
to identify the fields associated with the axial vector mesons
though, which, for example, stops us testing vector meson
dominance in the Weinberg sum rules \cite{Son}. Nevertheless we
compare the vector meson spectrum to that of the ${\cal N}=2$
theory provided by a pure AdS background.

\section{Chiral Dynamics}

QCD with $N_f$ massless quarks has a chiral $SU(N_f)_L\times
SU(N_f)_R$ global symmetry. When asymptotic freedom drives the
coupling strong it is believed that this symmetry is broken by a
quark bilinear condensate to the vector $SU(N_f)$ subgroup. The
symmetry breaking produces $N_f^2-1$ Goldstone bosons ($N_f^2$ at
large $N_c$ where instanton effects are suppressed) which are
associated with the pion multiplet in nature. Since the quarks
have small current masses the pions are only pseudo-Goldstone
fields. The very low energy dynamics only involves these Goldstone
fields and can be completely described by a theory that realizes
the broken symmetry non-linearly \cite{Coleman, Georgi}. Such a
phenomenological theory is called a chiral lagrangian. We
introduce fields $\Pi^a$ (with $T^a$ the generators of the broken
$SU(N_f)$ group normalized such that Tr$T^aT^b = {1 \over 2}
\delta^{ab}$) 

\beq U= exp(2 i \Pi^a T^a/f_\Pi), \hspace{1cm} 
U \rightarrow L^\dagger U R  \eeq 
where $f_\Pi$ is the pion decay constant
and where the chiral symmetry transformation properties are shown.
Remembering that the quark mass matrix transforms as $M
\rightarrow  L^\dagger M R$ we can then write a lagrangian at
leading order in both a derivative expansion and an expansion in
$M/f_\Pi$

\beq {\cal L} = { f_\Pi^2 \over 4} Tr \partial^\mu U \partial_\mu
U^\dagger + \nu^3 Tr M U^\dagger + \nu^3 Tr U M^\dagger + ....
\eeq expanding $U$ to find the leading terms for the pion fields
gives \beq \label{cl} {\cal L} = 2 N_f \nu^3 m + {1 \over 2}
(\partial^\mu \Pi^a)^2 - {1 \over 2} { 4 \nu^3 m \over f_\Pi^2}
\Pi^{a2} \eeq

The couplings $f_\Pi$ etc must be found phenomenologically in the
low energy theory but are in principle predictions of the full
high energy QCD theory. Since $M$ is a source for the quark
bilinear condensate the coupling $\nu$ is related to the value of
the quark condensate \beq \label{gmor} \nu^3 = { 1 \over 2}
\langle \bar{q} q \rangle \eeq Note the pion mass can then be
written as the Gell-Mann-Oakes-Renner relation $ m_\Pi^2 = {2 m_q
\langle \bar{q} q \rangle \over f_\pi^2}$ \cite{Gell}.

The interaction terms in the low energy theory at next order in
the chiral expansion have been written down by Gasser and
Leutwyler \cite{Gasser} and can be parameterized as

\beq \label{gl} \begin{array}{ccl} {\cal L} &= & L_1 tr (D^\mu U
D_\mu U^\dagger)^2 + L_2 tr ((D^\mu U D^\nu U^\dagger) (D_\mu U
D_\nu U^\dagger) ) + L_3 tr(D^\mu U D_\mu U^\dagger D^\nu U D_\nu
U^\dagger) \\&&+ L_4 tr (D^\mu U D_\mu U^\dagger) tr (M^\dagger U
+ M U^\dagger) + L_5 tr (D^\mu U D_\mu U^\dagger)(M^\dagger U + M
U^\dagger) \\&& + L_6(M^\dagger U + M U^\dagger)^2 + L_7 tr
(M^\dagger U - M U^\dagger) + L_8 tr(M^\dagger U M^\dagger U + M
U^\dagger M U^\dagger)
\\&& + i L_9 tr (F_{\mu \nu}^R D^\mu U D^\nu U^\dagger + F_{\mu
\nu}^L D^\mu U D^\nu U^\dagger) + L_{10} tr(U^\dagger F^R_{\mu
\nu} U F^{L\mu\nu}) + L_{11} tr(D^2U D^2 U^\dagger) \\&&+ L_{12}
tr(M^\dagger D^2 U + M D^2 U^\dagger)\end{array}\eeq

We will be interested in showing that the pions of our gravity
construction conform to this structure. We will also estimate some
of these coefficients below using the AdS/CFT Correspondence
method.

\subsection{Naive Dimensional Analysis}

A simple set of rules for estimating the size of chiral Lagrangian
couplings has been devised \cite{Georgi2}. In QCD there are two
scales - $\Lambda$ the strong coupling scale generated by the
running coupling, and the pion decay constant, $f_\pi$, which is
approximately

\beq f_\Pi^2 \sim {N \over (4 \pi)^2} \Lambda\eeq Naive
dimensional analysis says that one should give all chiral
lagrangian terms a common coefficient of $\Lambda^2 f_\Pi^2$ with
any occurrences  of $M$ or $D^\mu$ being dimensionally balanced by
a factor of $\Lambda$. Note that the pion fields enter in $U$
dimensionally balanced by $f_\Pi$. For example the $L_i$
coefficients are predicted to be of order $1/16 \pi^2$ using these
rules which reasonably matches their physical values.

One of our goals in this paper is to test this naive power
counting in the strongly coupled gauge theory for which we have a
gravitational dual. We will find reasonable agreement below.

\section{The Brane Construction and Gravity Dual}

AdS/CFT Correspondence type duals are obtained by equating the
physics on a stack of $N$ coincident, flat D3 branes and the
supergravity background dynamically generated by the D3 branes'
tension. We will consider the non-supersymmetric deformed AdS geometry
originally constructed in \cite{CM}. This geometry is dual to the
${\mathcal N}=4$ super Yang-Mills theory

\beq {\cal L} = { 1 \over g^2_{YM}} \left[{1 \over 4} Tr F^{\mu
\nu} F_{\mu \nu} + ...\right] \eeq (with $g^2_{YM}= 2 \pi g_s$)
deformed by the presence of a vacuum expectation value for an
R-singlet operator with dimension four (such as $Tr F^{\mu
\nu}F_{\mu \nu}$). The supergravity background has a dilaton and
$S^5$ volume factor depending on the radial direction. The
geometry has a naked singularity in the interior which we loosely
expect to correspond to the presence of the central stack of D3
branes.  Whether this geometry actually describes the stable
non-supersymmetric vacuum of a field theory is not well understood
\cite{CM}.  This is not so important from our point of view though
since the geometry is a well defined gravity description of a
non-supersymmetric gauge configuration. We will just study the
behaviour of quarks in that background.

The geometry in {\it Einstein frame} is given by

\beq ds^2 = H^{-1/2} \left( { w^4 + b^4 \over w^4-b^4}
\right)^{\delta/4} dx_{4}^2 + H^{1/2} \left( {w^4 + b^4 \over w^4-
b^4}\right)^{(2-\delta)/4} {w^4 - b^4 \over w^4 } \sum_{i=1}^6
dw_i^2, \eeq where

\beq H =  \left(  { w^4
+ b^4 \over w^4 - b^4}\right)^{\delta} - 1
\eeq
and the dilaton and four-form become

\beq e^{2 \phi} = e^{2 \phi_0} \left( { w^4 + b^4 \over w^4 - b^4}
\right)^{\Delta}, \hspace{1cm} C_{(4)} = - {1 \over 4} H^{-1} dt
\wedge dx \wedge dy \wedge dz. \eeq There are formally two free
parameters, $R$ and $b$, since \beq \delta = {R^{4} \over 2 b^4},
\hspace{1cm} \Delta^2 = 10 - \delta^2 \eeq

As usual in the AdS/CFT Correspondence the $w$ directions have the
conformal scaling properties of energy scale in the field theory.
Thus $b$ is the only object that breaks the conformal (and super)
symmetry of the gauge theory. We define an associated mass scale

\beq \Lambda_b = { b \over 2 \pi \alpha'}\eeq which is the mass of
a string of length $b$. Since $\Lambda_b$ is the scale of
conformal symmetry breaking and the theory is strongly coupled at
that scale we expect the dynamical strong coupling scale of the
theory, $\Lambda$, to lie close to $\Lambda_b$. We will hence
forth loosely associate the two. In fact we are not interested in
changing the scale $\Lambda_b$ since it is the only mass scale so
we can set $b=1$ below. First though lets consider the $R$
dependence of the solution.

The parameter $R$ determines $g^2_{YM}N$ in the field theory as
usual in the Correspondence ($R^2 = \sqrt{4 \pi g_s N} \alpha'$).
We find it easiest to track the $R$ dependence by writing $w$ and
$b$ in units of $R$ so that $\delta = 1/2 b^4$. This means that
our fundamental energy scale $\Lambda_b$ scales with $R$ so we
should express masses as a ratio of

\beq \Lambda_b = {R b \over 2 \pi \alpha'}\eeq Now we can set
$b=1$ and the metric we will use is

\beq ds^2 = H^{-1/2} \left( { w^4 + 1 \over w^4-1}
\right)^{\delta/4} dx_{4}^2 + R^2 H^{1/2} \left( {w^4 + 1 \over
w^4- 1}\right)^{(2-\delta)/4} {w^4 - 1 \over w^4 } \sum_{i=1}^6
dw_i^2, \eeq where

\beq H =  \left(  { w^4 + 1 \over w^4 - 1}\right)^{\delta} - 1 ,
\hspace{1cm} e^{2 \phi} = e^{2 \phi_0} \left( { w^4 + 1 \over w^4
- 1} \right)^{\Delta} \eeq \vspace{0.3cm}

We will now introduce one flavour of quark via a D7 brane probe in
the geometry. The D7 brane lies in the $x_{4}$ directions and
$w_1-w_4$ (it is convenient to define a coordinate $\rho$ such
that $\sum_{i=1}^4 dw_i^2 = d\rho^2 + \rho^2 d\Omega_3^2$). This
configuration in pure AdS preserves ${\cal N}=2$ supersymmetry on
the D3 world volume so corresponds to introducing a quark
hypermultiplet $Q$ and $\tilde{Q}$. ${\cal N}=2$ supersymmetry
will ensure there is a superpotential coupling to one of the three
adjoint chiral multiplets $A$ of the original ${\cal N}=4$ theory
($W = \tilde{Q} A Q$).  In the non-supersymmetric field theory we
expect the scalar fields to gain masses of order the supersymmetry
breaking scale. The remaining fermionic terms in the gauge theory
are of the form

\beq {\cal L} = {1 \over g^2_{YM}} \left[ \bar{q}~ \slash
\hspace{-0.3cm} D ~q + ... \right] \eeq where one should note that
$g_s$ enters the normalization as for the gauge fields (this is
not the standard normalization for comparison with eg eq
(\ref{gmor}) above).

The Dirac Born Infeld action for the probe is

\beq S_{D7} = - { 1 \over (2 \pi)^7 \alpha^{'4} g_s} \int d^8 \xi
e^{\phi} \sqrt{- {\rm det}(P[G_{ab}])} \eeq where P indicates the
pullback of the space-time metric onto the D7 world volume. This
action will determine how the D7 lies in the remaining $w_5-w_6$
directions. We will use a complex coordinate in this plane \beq
\Phi=w_{6}+i w_{5}=\sigma e^{i\theta} \eeq

Asymptotically where the geometry returns to AdS the resulting
Euler Lagrange equation is

\beq {d \over d \rho} \left[ \rho^3 {d \Phi \over d \rho}\right]
=0 \eeq and has solutions \beq \Phi = m + {c \over \rho^2}\eeq The two
integration constants correspond to a mass and vev for the quark
bilinear $\bar{q} q$ with

\beq m_q = { m R\over 2 \pi \alpha'}, \hspace{1cm}\bar{q} q = {c R
\over 2 \pi \alpha'} \eeq Note that we are writing $\Phi$ in units
of $R$ and hence it has the same $N$ scaling as the parameter $b$.
These measures of the quark mass and condensate when expressed as
multiples of $\Lambda_b$ do not scale with $N$.

In the massless limit where the D7 brane lives at $\Phi=0$ there
is a U(1) symmetry acting in the $\Phi$ plane. Clearly this
corresponds to an angular rotation on $\bar{q} q$ and is hence the
$U(1)_A$ symmetry of the quarks. In fact this symmetry is also
part of the isometries of the space transverse to the central D3
brane stack and is thus part of the $SO(6)_R$ symmetry of the
gauge background. This reflects the presence of the UV
superpotential term $W = \tilde{Q} A Q$ which mixes the axial and
R symmetries. The $A$ field is hopefully somewhat massive in this
configuration (of order $\Lambda_b$) but the symmetry is
nevertheless this mixture.

\section{Chiral Symmetry Breaking}

Let us begin by considering the massless quark limit where there
is a good $U(1)_A$ symmetry. Asymptotically in the UV the D7 brane
lies at $\Phi=0$. The equation of motion that determines how it
lies in the interior is given by

\beq \label{eommc}{ d \over d \rho} \left[ {e^{\phi}  { \cal
G}(\rho,\Phi) \over \sqrt{ 1 + |\partial_\rho \Phi|^2}  }
(\partial_\rho \Phi)\right] - \sqrt{ 1 + |\partial_\rho \Phi|^2} {
d \over d \bar{\Phi}} \left[ e^{\phi} { \cal G}(\rho,\Phi) \right]
= 0. \eeq where

\beq {\cal G}(\rho,\Phi) =  \rho^3 {( (\rho^2 + |\Phi|^{2})^2 + 1)
( (\rho^2 + |\Phi|^{2})^2 - 1) \over (\rho^2 + |\Phi|^{2})^4}.
\eeq The final terms in the equation of motion is a ``potential"
like term that is evaluated to be

\beq { d \over d \bar{\Phi}} \left[ e^{\phi} { \cal G}(\rho,\Phi)
\right] = { 4 \rho^3 \Phi \over (\rho^2 + |\Phi|^2)^5} \left( {(
(\rho^2 + |\Phi|^2)^2 + 1) \over ( (\rho^2 + |\Phi|^2)^2 - 1)}
\right)^{\Delta/2} ( 2  - \Delta (\rho^2 + |\Phi|^2)^2). \eeq

The equation of motion has an explicit $\Phi \rightarrow e^{i
\alpha} \Phi$ symmetry. In the massless case this is the $U(1)_A$
symmetry on the quarks. The solutions to the equation of motion in
the interior simplify to the D7 brane sitting at a fixed angle in
the plane, $\theta$, with the radial behaviour given by setting
$\Phi = \sigma$ real in the equation of motion.

There are two regular solutions of the equation of motion with UV
boundary condition $\sigma = c/\rho^2$, shown in Figure 1. They
have $c= \pm 1.86$ showing that the solution indeed prefers the
formation of a chiral symmetry breaking condensate. Note that in
the IR (small $\rho$) these solutions are $\sigma=$ constant, so
they can also be found by numerically solving up from the IR and
matching to the UV boundary conditions. This is the simplest way
to find solutions since any IR boundary condition of this sort
produces a regular flow and it is then not necessary to tune onto
the regular flows. The two solutions are just two opposing points
on a circle in $\Phi$ verifying that there is indeed a set of
solutions with the same radial behaviour at each value of $\Phi$.
This circle of degenerate solutions is the vacuum manifold.

\begin{figure}[!h]
\begin{center}
\includegraphics[height=7cm,clip=true,keepaspectratio=true]{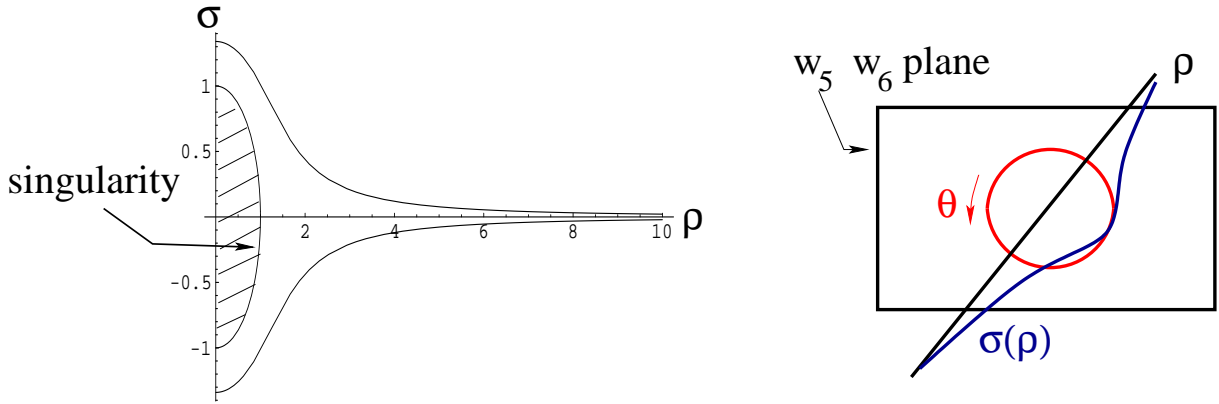}
\caption{Solutions of (\ref{eommc}) for the position of the D7
brane showing chiral symmetry breaking in the massless limit and a
sketch of the D7s position in the space.}\label{fig1}
\end{center}
\end{figure}

It is interesting to see how the chiral symmetry breaking
manifests itself geometrically. The solution for the position of
the D7 brane is sketched in Figure 1. Asymptotically it lies at
the origin of the plane but in the interior the D7 brane is
repelled by the core of the geometry forcing it out into the plane
where it breaks the U(1) symmetry.

\subsection{Massive Case}

It is straightforward to also introduce a mass into the UV
boundary conditions for $\Phi$. Let us first look at solutions of
the form $\theta=0$ and $\sigma = m + c/\rho^2$ in the UV. These
will turn out to describe the true vacuum of the theory. The
regular solutions have $\sigma$ tend to a constant in the IR so
are easily found by flowing up from the IR. The mass and
condensate can then be read from the solution in the UV. We plot a
few of these solutions in Figure 2. This procedure also allows us
to determine the mass dependence of the condensate and we plot
that also in figure 2.

\begin{figure}[!h]
\begin{center}
\includegraphics[height=4.85cm,clip=true,keepaspectratio=true]{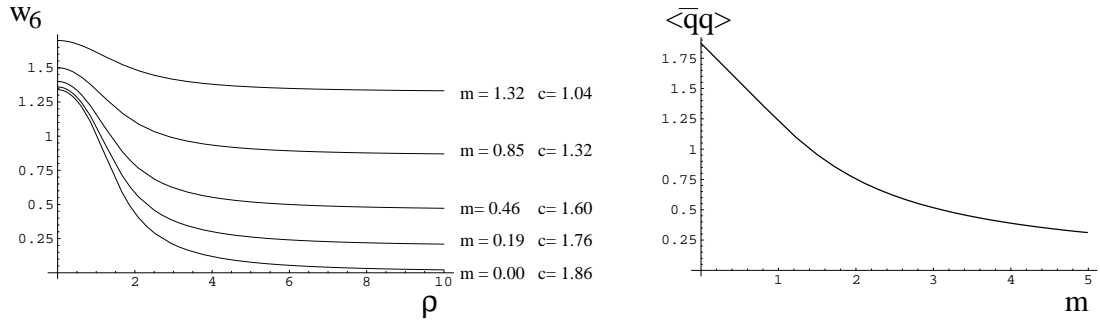}
\caption{Solutions for the $w_6$ flow when $w_5=0$ showing
the dependence of the condensate on the quark mass.}\label{cVsm}
\end{center}
\end{figure}

It is also interesting to study the vacua with a real mass term
but with a phase on the condensate - these will correspond to the
vacua around the circle in $\theta$ in the $m=0$ limit and we
expect their energy to be lifted relative to the true vacuum when
the mass is non-zero. To study these we use boundary conditions on
the UV fields

\beq w_5 = m + {c \over \rho^2} \cos \theta, \hspace{1cm} w_6 ={c
\over \rho^2} \sin \theta \eeq In general we are looking for a
regular solution for both $w_5, w_6$ and this is very hard to
achieve numerically. As an example of evidence for the existence
of these solutions though we show in Figure 3 some solutions for
$w_5, w_6$ as a function of $\rho$ for the case $m=0.1$,
$\theta=90^o$. The solutions are plotted with asymptotic UV
boundary conditions with two different values of the condensate
parameter $c$. Both $w_5, w_6$ fields change behaviour in this
range suggesting there may indeed be a regular solution in
between. We have not been able to pin down the solution to greater
accuracy than this though.

\begin{figure}[!h]
\begin{center}
\includegraphics[height=5cm,clip=true,keepaspectratio=true]{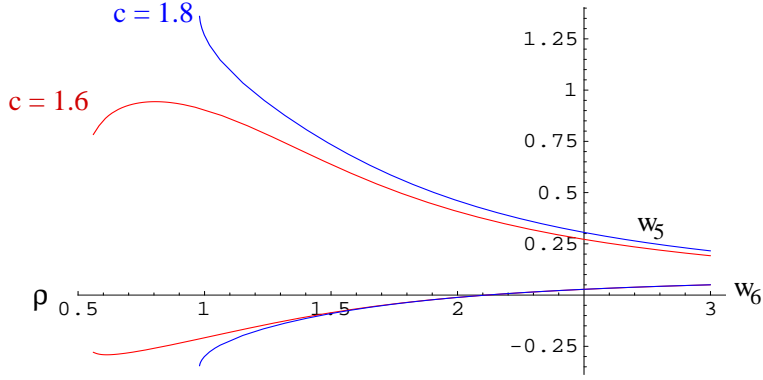}
\caption{$w_5, w_6$ flows for $m=0.1$, $\theta=90^o$ showing the
possibility of a regular flow between these values of $c$.}
\end{center}
\end{figure}

More easily we can study the two solutions of the equation of motion with $\Phi$ taken real.
These correspond to the cases of
 $\pi=0^o$ and $180^o$. The two solutions have the same value of m in the UV
 but opposite c. Example solutions for a mass of 0.46 are plotted
 in Figure 4.

\begin{figure}[!h]
\begin{center}
\includegraphics[height=5cm,clip=true,keepaspectratio=true]{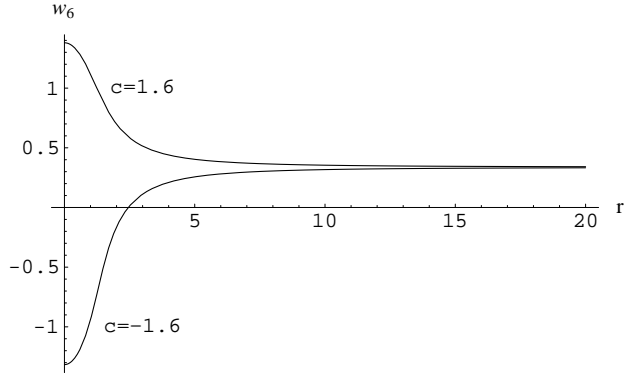}
\caption{$w_6$ flows for m=0.46 with positive and negative
condensate solutions }\label{fig2}
\end{center}
\end{figure}

We can now begin to determine the chiral Lagrangian parameters
predicted by this model. We can calculate the vacuum energy as a
function of the quark mass for the positive and negative
condensate solutions. The vacuum energy is given by the DBI action

\beq S_{D7} = - { R^4 \over (2 \pi)^7 \alpha^{'4} g_s} \int d^8 \xi
e^{\phi}  {\cal G}(\rho, \Phi) \sqrt{ 1 + (\partial_\rho \Phi)^2}
\eeq The angular integral over the $S^3$ the D7 wraps gives $2
\pi^2$. The resulting four dimensional cosmological term,
$\Omega^4$, should be normalized by $\Lambda_b^4$

\beq \label{om} {\Omega^4 \over \Lambda_b^4} = { 1 \over 2 \pi
g_s} {\cal I}_0, \hspace{1cm} {\cal I}_0 = {1 \over 2} \int d \rho
e^{\phi} {\cal G}(\rho, \Phi) \sqrt{ 1 + (\partial_\rho
\Phi)^2}\eeq where the integral ${\cal I}_0$ is over the solutions
described above. Note that the factor of $g_s$ shows us that the
vacuum energy scales as $N$ in the large $N$ limit with $g_s N$
kept constant. This is consistent with expectations for the vacuum
energy contributions from fundamental quarks.

Asymptotically $\Omega^4 \simeq \rho_{UV}^4$ indicating the
expected UV divergence of the vacuum energy. This will be present
for all our configurations no matter the quark mass or condensate.
We are interested in the corrections due to the low energy chiral
dynamics (those in eq(\ref{cl}))  so subtract the massless result
for $\Omega^4$ from the vacuum energy for each configuration. We
show numerical results for $\Omega^4 / \Lambda_b^4$ in Figure 5.

\begin{figure}[!h]
\begin{center}
\includegraphics[height=5cm,clip=true,keepaspectratio=true]{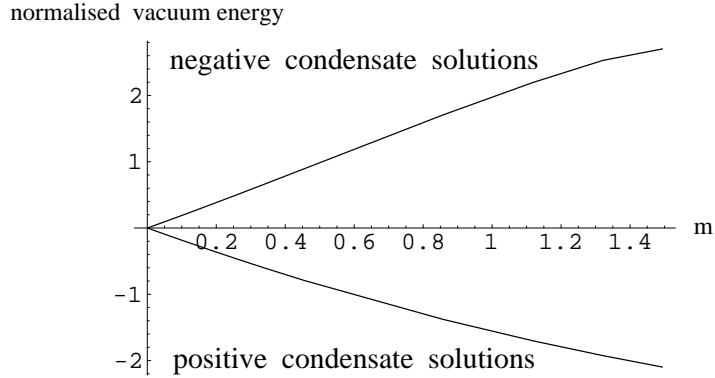}
\caption{Vacuum energy, with that at $m=0$ subtracted, for
$\theta=0^o$ and $180^O$ showing a lower energy for the case
 where the mass and condensate are both positive. }\label{Mvsvacuum}
\end{center}
\end{figure}

The positive condensate solutions are energetically favourable.
This is because the potential for the quark condensate has a
tilted wine bottle shape of the form shown in Figure 6. There we
sketch a potential of the form $V = \alpha |\Phi|^4 - \beta^2
|\Phi|^2 + m Re(\Phi)$ - as $m$ increases the potential tilts
giving a single true vacuum. In this simple model there is a
critical value of $m$ where the central ``hump" disappears and
there is a single unique vacuum. We have numerically looked for
such a solution in the D7 brane case. As the boundary condition on
$m \rightarrow 1.5$ the solution for the meta-stable vacuum (of
the type shown in Figure 4) has a singular derivative as it passes
through the $w_6$ axis and we loose numerical control. This may
well be an indication of the absence of such solutions for larger
$m$ corresponding to the loss of the hump in the naive model.

\begin{figure}[!h]
\begin{center}
\includegraphics[height=5cm,clip=true,keepaspectratio=true]{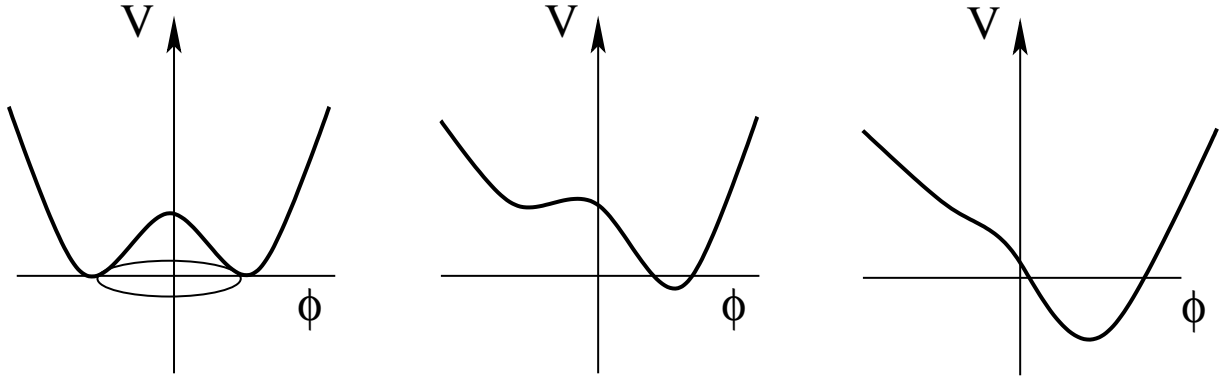}
\caption{Sketch of a simple Higgs potential with explicit symmetry
breaking term $m Re(\Phi)$, showing spontaneous symmetry breaking
potential as $m$ increases.}\label{fig3}
\end{center}
\end{figure}

Finally we can use the vacuum energy to compute $\nu^3$ in eq
(\ref{cl}) and to provide an alternative identification of the
quark condensate in the model. The chiral lagrangian predicts that
the vacuum energy will be given by

\beq \Omega^4 = 2 \nu^3 m = { \langle \bar{q} q \rangle m} \eeq

Before extracting $\langle \bar{q} q \rangle$ in this way we must
normalize the quark fields in the standard way appropriate for
these equations. This removes the factor of $1/2\pi g_s$ in eq
(\ref{om}) and we then extract the condensate from the slope of
the vacuum energy vs mass plot. We find numerically

\beq \langle \bar{q} q \rangle = 1.86\eeq which almost precisely
matches the value of $c$ in the $m \rightarrow 0$ limit above. The
fact that $c$ corresponds to the condensate of the canonically
normalized fields explains why it does not scale with $N$. This
identification confirms $c$ is the condensate or equally the form
of the Gell-Mann-Oakes-Renner relation.

\section{Pions and Their Interactions}

We will now turn to studying the Goldstone bosons of the chiral
symmetry breaking we have observed above. So far we have
considered the dynamical breaking of the $U(1)_A$ symmetry of a
single flavour of quark. The Goldstone boson will therefore be the
analogue of the $\eta'$ in QCD, which becomes degenerate with the
QCD pions in the limits where the quark masses vanish and at large
$N$ where the anomaly is suppressed. We will loosely refer to our
state as a pion. We will shortly consider  extending this
construction to non-abelian axial symmetries.

\subsection{The Pion Mass and $f_\pi$}

\begin{figure}[!h]
\begin{center}
\includegraphics[height=5cm,clip=true,keepaspectratio=true]{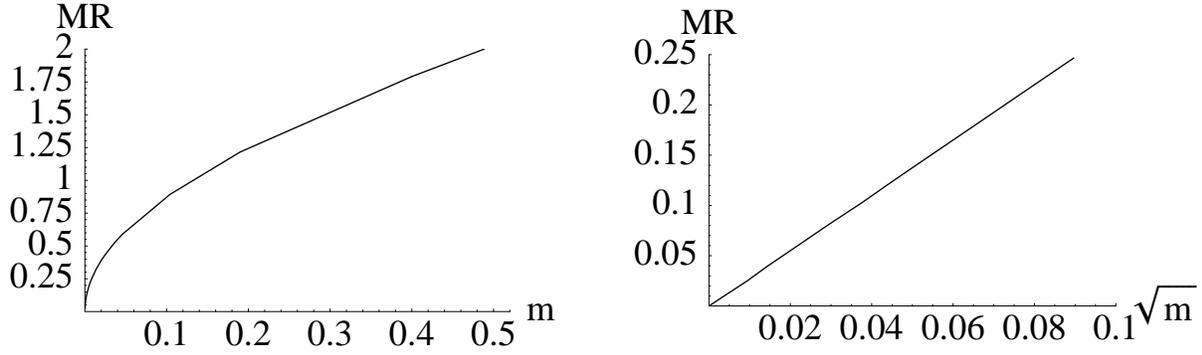}
\caption{Quark mass versus meson mass showing massless pion at $m=0$.
Also shown is the square root behaviour of the relationship }\label{mVsM}
\end{center}
\end{figure}

The pion in the one flavour case will be a bound state of a quark
and an anti-quark and hence will be described by the D7 world
volume field $\Phi$. The Goldstone fluctuation corresponds to
oscillations of the field along the vacuum manifold - in the brane
language this will correspond to fluctuations of the brane in the
direction of the possible set of solutions for its background
position. In the chiral limit this corresponds to fluctuations in
the angular $\theta$ direction in the $w_5-w_6$ plane. To leading
order, if we have a background configuration for the D7 brane
described by a particular real solution of (\ref{eommc}),
$\sigma_0$, we can look at small fluctuations in the $\theta$
angular direction

\beq \theta(\rho,x) = f(\rho) \sin kx \eeq  We look for solutions
where $k^2 = M^2$. The action for these fluctuations to quadratic
order is given by expanding the DBI action (note that the metric
is independent of the angle $\theta$)

\beq \label{act} S = { 2 \pi^2 \over (2 \pi)^7 \alpha^{'4} g_s}
\int d\rho d^4 x e^\phi R^4 {\cal G} \sqrt{1 + (\partial_\rho
\sigma_0)^2} \left[ 1 + {1 \over 2} { g^{\rho \rho} g_{\theta
\theta} (\partial_\rho \theta)^2 \over (1 + (\partial_\rho
\sigma_0)^2)} + {1 \over 2}\frac{g^{\mu \nu} g_{\theta \theta}
(\partial_\mu \theta) (\partial_\nu \theta)}{(1 + (\partial_\rho
\sigma_0)^2)} \right] \eeq Note that the DBI action only produces
terms with derivatives of $\theta$ - we can though obtain
potential terms in the pion field when those derivatives are
$\rho$ derivatives that act on $f(\rho)$. We find the resulting
equation of motion for $f$

\beq { d \over d \rho} \left[ {e^\phi {\cal G} \over \sqrt{1 +
(\partial_\rho \sigma_0)^2}} \sigma_0^2 (\partial_\rho f) \right]
+\frac{ R^2 M^2 e^\phi {\cal G}}{ \sqrt{1 + (\partial_\rho
\sigma_0)^2}} H \left( { (\rho^2 + \sigma_0^2)^2 + 1 \over (\rho^2
+ \sigma_0^2)^2 - 1} \right)^{(1-\delta)/2} { (\rho^2 +
\sigma_0^2)^2 - 1 \over (\rho^2 + \sigma_0^2)^2}~ \sigma_0^2 f =0
\eeq This equation can be numerically solved for $MR$ as a
function of $m$ using the UV boundary condition $f = 1/\rho^2$
(reflecting the fact that the pion has the UV scaling dimension of
$\bar{q} q$) and seeking regular solutions for $f$. We plot the
result (in which $m$ determines the $\sigma_0$ flow) in Figure
\ref{mVsM}. We indeed find a massless pion at $m=0$ in accordance
with Goldstone's theorem. Note that in the case where the pion
mass vanishes the equation of motion is just the first term above
and when substituted back into (\ref{act}), after integration by
parts, explicitly makes the $(\partial_\rho \theta)^2$ term
vanish. In the chiral limit if one worked to higher order all
higher order terms involving $(\partial_\rho \theta)^n$ would also
vanish directly. This demonstrates that the vacuum manifold is a
truly flat direction. Below we will only look at terms at higher
order involving the spatial derivatives $\partial_\mu$.

In figure \ref{mVsM} we also plot  $MR$ vs $\sqrt{m}$ for small
$m$ - there is a good linear fit matching the expectations for the
pion mass dependence on $m$ in the chiral lagrangian (\ref{cl}).
We will write

\beq {MR } = \kappa \sqrt{ m }\eeq where we have numerically
determined $\kappa = 2.75$.

Note that the appropriate four dimensional, low energy action can
be found by writing $\theta = 2 \pi \alpha' f(\rho) \Pi(x)$ and
substituting the equation of motion for $f(\rho)$ back into the
action (\ref{act}), after integrating by parts. This gives

\beq \label{norm} {\cal L}  =  ~ {\frac{R^6}{16 \pi^3\alpha'
g_s}}~{\cal I}_1 \left[ { 1 \over 2} (
\partial^\mu \Pi)^2  - { 1 \over 2} M^2 \Pi^2 + ... \right]
\eeq where \beq {\cal I}_1 = \int d \rho f(\rho)^2 \frac{ e^\phi
{\cal G}}{ \sqrt{(1 + (\partial_\rho \sigma_0)^2)}} H \left( {
(\rho^2 + \sigma_0^2)^2 + 1 \over (\rho^2 + \sigma_0^2)^2 - 1}
\right)^{(1-\delta)/2} { (\rho^2 + \sigma_0^2)^2 - 1 \over (\rho^2
+ \sigma_0^2)^2}~\sigma^2_0\eeq The coefficient at the front of
(\ref{norm}) should be absorbed into the normalization of $\Pi$ to
make the kinetic term canonical. We must be careful to include
this normalization below. Note also that ${\cal I}_1$ is only
defined upto the normalization of $f$ which is free in the
linearized equation of motion - this factor will cancel when $\Pi$
is canonically normalized in all quantities below.

The chiral Lagrangian predicts that

\beq M^2_{\pi} = { 4 \nu^3 m \over f_\pi^2} \eeq Since we have
extracted $\nu^3$ from the vacuum energy above we can now
determine $f_\pi$. With the supersymmetric normalization of fields
we found numerically\beq {\nu^3 \over \Lambda_b^3} = { 1 \over 4
\pi g_s } c\eeq Writing our result for the pion mass in units of
$\Lambda_b$ gives \beq {M^2 \over \Lambda_b^2} = \kappa^2
 { m_q \over \Lambda_b} {
\pi \over g_s N} \eeq and hence we find for $f_\pi$

\begin{eqnarray}
{f_\pi^2 \over \Lambda_b^2}&=& { N \over  \pi^2
\kappa^2}c\nonumber\\
&=&0.246 \frac{ N}{\pi^2}
\end{eqnarray}
which has the expected dependence on $N$ and is suppressed
relative to $\Lambda_b$ by roughly a factor of $4 \pi^2$. This
seems to match well with the naive estimate of section 2.

\subsection{Higher Order Interactions}

It is also interesting to study higher order interaction terms
such as those in (\ref{gl}). To do this we can expand the DBI
action beyond linearized order in $\theta$ and use the linearized
solution for $f(\rho)$.

In the chiral limit the identification of the pion field with
fluctuations of the brane in the $\theta$ direction is rigorous.
However, when there is a quark mass present the vacuum will be
distorted from a true circle centered on the origin in $\Phi$ and
this procedure does not correctly identify the pseudo-Goldstone
boson (the angular direction will have some of the massive
``higgs" like mode mixed in). In principle to find the pseudo
Goldstone boson in the massive case one would need to know how to
separate the function $\sigma_0(r)$ into pieces that represent the
running of the mass and the condensate separately. The pseudo
Goldstone would correspond to angular fluctuations about just the
condensate piece. In other words simply changing the phase on
$\sigma_0$ changes the phase on the mass which one does not want
to do. We have not been able to resolve this so will simply study
the massless chiral limit here.

As mentioned above in the chiral limit any terms in the action
involving $(\partial_\rho \theta)^n$ vanish by the equations of
motion so we will neglect those here (in fact we have checked they
explicitly vanish when evaluated on our solution for $f(\rho)$ in
this limit). Now we can study the Gasser-Leutwyler coefficients
$L_{1}$, $L_{2}$, $L_{3}$. In the chiral Lagrangian, these appear
in the form:

\beq L_{1}tr(\partial^{\mu}U\partial_{\mu}U^{\dagger})^{2}+
L_{2}tr((\partial^{\mu}U\partial^{\nu}U^{\dagger})(\partial_{\mu}U\partial_{\nu}U^{\dagger}))+
L_{3}tr(\partial^{\mu}U\partial_{\mu}U^{\dagger}\partial^{\nu}U\partial_{\nu}U^{\dagger})
\eeq

Since we only have an abelian U(1) symmetry breaking pattern, we
can not differentiate between these terms. When U is expanded,
each term has a factor of $(\partial^{\mu} \Pi)^4$ so we can't
extract any information about $L_{1}$, $L_{2}$ or $L_{3}$
separately. It is still interesting to extract the coefficient of
the $(\partial^{\mu} \Pi)^4$ term. This term in the action is

\beq {\cal L} = { 2 \pi^2 R^4 \over (2 \pi)^7 g_s \alpha^{'4}}
\int d\rho e^\phi {\cal G} \sqrt{1 + (\partial_\rho \sigma_0)^2}
\left[ - {1 \over 4} {(g_{\theta\theta} g^{\mu \nu})^2 \over (1 +
(\partial_\rho \sigma_0)^2)} (\partial_\mu \theta)^4 \right] \eeq
Rewriting this in terms of the field $\Pi$ gives

\beq {\cal L} = - {R^8 \over 16 \pi g_s} {\cal I}_2 (\eta^{\mu
\nu}\partial_\mu \Pi \partial_\nu \Pi)^2 \eeq where \beq {\cal
I}_2 = \int d\rho e^\phi {{\cal G} f(\rho)^4\sigma_{0}^4 \over
(1+(\partial_\rho
\sigma_0)^2)^{3/2}}\left[H^2\left(\frac{\omega^4+1}{\omega^4-1}
\right)^\frac{1}{2}\left(\frac{\omega^4-1}{\omega^4}\right)^2\right]
\eeq

We must remember to normalize the pion field so the kinetic term
in (\ref{norm}) is canonically normalized from which we find the
coefficient of the $(\partial_\mu \Pi)^4$ term (which we loosely
write as $4 L/f_\pi^4$) is given by

\beq {L \over f_\pi^4} = { 16 \pi^5 \alpha^{'4} g_s \over
R^4}\frac{{\cal I}_2}{{\cal I}_1^2}\eeq or using the expression
above for $f_\pi$

\beq L = {(g^2_{YM} N) N  c^2 \over \kappa^4 2 \pi^4}\frac{{\cal
I}_2}{{\cal I}_1^2} \eeq

This has the same large $N$ scaling (scaling as $f_\Pi^2$) as the
naive dimensional analysis estimate. Numerically we find ${\cal
I}_2 /{\cal I}_1^2 =0.29$. If we take $g^2_{YM} N = (4 \pi)^2$
which seems a reasonable strong coupling value then we find
$L\simeq 0.016 N$. This is about a factor of 2-3 bigger than $N/16
\pi^2 = 0.006N$ and hence broadly consistent with naive
dimensional analysis.

\subsection{Non-Abelian case}

Naively we can introduce extra flavours of quarks by including
additional probe D7 branes and using the non-abelian DBI action
\cite{nadbi}. As mentioned in the introduction we do not generate
a U(N) axial symmetry because of the superpotential term
$\tilde{Q}AQ$ in the UV theory. The adjoint scalar though is
expected to have a mass of order $\Lambda$ and we might expect the
theory to have an accidental chiral symmetry. The non-abelian DBI
action requires us to treat the scalar field $\Phi$ as a matrix
which for the pions will require us to write

\beq \Phi= \sigma_0 e^{i \Pi^a T^a} \eeq

Where $T^a$ are the broken generators of the Lie group. With $N_f$
D7 branes this is $U(N_f)$.  The DBI action acquires a trace over
the flavour indices. As pointed out in \cite{Mateos2} if this were
the only change then we would find $N_f^2$ copies of the action we
had above and hence $N_f^2$ massless pions. The DBI action also
contains terms though that generate the usual commutator scalar
interactions ($tr[\phi_i, \phi_j]^2$) - these do not though
generate a mass for the pions. The explicit chiral symmetry
breaking appears therefore to be quite weak in its effects. To
pursue this issue further we can study whether the interactions of
the pions, neglecting the commutator terms, coincide with
expectations from chiral symmetry.

For simplicity we will work in the case with just two flavours and
look at just the interactions of two pions, $\Pi_1$ and $\Pi_2$.
This means that we should write: \beq \Phi = \sigma_0 Exp \left[ i
(2 \pi \alpha') f(\rho)\left(\begin{array}{cc}
0 & \Pi_1(x)+I\Pi_2(x) \\
 \Pi_1(x)-I\Pi_2(x) & 0
\end{array}
\right) \right]\eeq Note that in the chiral limit we can drop all
terms in the DBI action involving $\rho$ derivatives since they
will cancel by the equation of motion as discussed above. The
resulting action then takes the form of a chiral lagrangian with
spatial derivative acting on the field $U$. We find the lagrangian
for the two pion fields takes the form

\beq \begin{array}{ccl} {\cal L} & = & {\cal
A}\left[(\partial_{\mu}\Pi_{1}(x))^2+(\partial_{\mu}\Pi_{2}(x))^2\right]\\
&&\\ && +  {1 \over 3} {\cal B} \left[2
\Pi_{1}(x)\Pi_{2}(x)(\partial_{\mu}\Pi_{1}(x)\partial^{\mu}\Pi_{2}(x))
-
\Pi_1^2(\partial_{\mu}\Pi_{2}(x))^2+\Pi_{2}^2(\partial_{\mu}\Pi_{1}(x))^2\right]
\end{array}
\eeq where \beq \begin{array}{ccc} {\cal A} & = & \frac{2\pi^2
R^4}{(2 \pi)^7 \alpha'^4 g_s} (2\pi\alpha'R)^2 {\cal
I}_1\\
&&\\
{\cal B} & = &  \frac{2\pi^2 R^4}{(2 \pi)^7 \alpha'^4 g_s} ( 2 \pi
\alpha')^4 R^2 {\cal I}_3\end{array}\eeq and \beq {\cal I}_3 =
\int e^\phi  {\cal G} H\left(\frac{w^4+1}{w^4-1}
\right)^\frac{1}{4}\frac{w^4-1}{w^4}\frac{\sigma_0^2
f^4}{\sqrt{(1+(\partial_\rho \sigma_0)^2)}} \eeq

Performing the equivalent expansion in the chiral lagrangian to
compare these terms we find:
\begin{eqnarray}
{\cal
L_\chi}&=&\frac{1}{2}((\partial_{\mu}\Pi_1(x))^2+(\partial_{\mu}\Pi_2(x))^2)\nonumber
+\frac{1}{3 f_{\Pi}^2}(\Pi_{1}(x)\Pi_{2}(x)(\partial_{\mu}\Pi_{1}
(x)\partial^{\mu}\Pi_{2}(x)))\\
&&-\frac{1}{6 f_{\Pi}^2}(\Pi_{1}
(x)^2(\partial_{\mu}\Pi_2(x))^2+\Pi_{2}(x)^2(\partial_{\mu}\Pi_1(x))^2)
\end{eqnarray}

Note that the relative sizes of the two new interaction terms
match between the DBI and chiral lagrangians (in both cases
because they result from the expansion of the same exponential
form).The two new interaction terms provide an additional
opportunity to calculate $f_\Pi$. Again, we canonically normalise
the kinetic term  and compare with the chiral lagrangian:
\begin{eqnarray}
\frac{f_{\pi}^2}{\Lambda_{b}^2}&=&\frac{{\cal I}_{1}^2}{{\cal
I}_{3}}
\frac{4AR^2}{6}\frac{(2\pi\alpha')^2}{R^2}\nonumber\\
&=&\frac{{\cal I}_{1}^2}{{\cal I}_{3}}\frac{N}{6\pi^2}\nonumber\\
&=&0.246\frac{N}{\pi^{2}}
\end{eqnarray}

Within our numerical accuracy this is the same answer as that
calculated in the abelian case from the mass term which gave the
value $\simeq \frac{N}{4\pi^2}$. The extra massless bosons indeed
seem to behave as Goldstone fields upto the commutator
interactions.

\section{Vector Mesons and Weakly Gauged Chiral Symmetries}

There is one additional bosonic field in the low energy DBI action
of the D7 brane - the gauge field partner of the scalar, $\Phi$,
discussed above. The field strength tensor of this gauge field
enters in the standard way as $2 \pi \alpha' F^{ab}$ in the square
root. The leading lagrangian for this field, in a background
configuration $\sigma_0$, is

\begin{eqnarray}
{\cal L} &=& {2 \pi^2 R^4 \over (2 \pi)^7 \alpha^{'4} g_s} \int d
\rho e^\phi {\cal G}\left( \sqrt{1 + (\partial_\rho \sigma_0)^2} H
\left(\frac{ (\rho^2 + \sigma_0^2)^2 - 1}{ (\rho^2 + \sigma_0^2)^2
+ 1}
\right)^\frac{1}{4} (2 \pi \alpha')^2 \frac{1}{4} F^{\mu\nu} F_{\mu\nu}\right.\\
&+&\left.\frac{1}{2 R^2}{1 \over \sqrt{1 + (\partial_\rho
\sigma_0)^2}}\left(\frac{(\rho^2 + \sigma_0^2)^2 - 1}{ (\rho^2 +
\sigma_0^2)^2 + 1} \right)^\frac{1}{2}\frac{(\rho^2 +
\sigma_0^2)^2}{(\rho^2 +
\sigma_0^2)^2-1}F^{\mu\rho}F_{\mu\rho}\right)
\end{eqnarray}
where $\mu$ and $\nu$ run over the space-time indices.  There is
an additional term that could be added onto the DBI action. This
is the Wess Zumino term which gives the coupling of the four-form
$C^{(4)}$ to the gauge fields. We do not include this because when
we calculate the equations of motion for the gauge fields, this
term is only relevant for the gauge fields with a vector index on
the $S^{3}$. We will only be interested in the states that carry
no SO(4) R-charge where $a,b$ take values in the four dimensional
space of the gauge theory. We write the gauge field as: \beq
A^{\mu}=g(\rho) \sin{(k x)} \epsilon^{\mu} \eeq

The equations of motion for the gauge field are given by:\beq e^\phi {\cal G}\sqrt{1+(\partial_{\rho}\sigma_0)^{2}}
M^2 R^2 g(\rho) H \left(\frac{w^4-1}{w^4+1}
\right)^\frac{1}{4}+\partial_\rho \left(\frac{e^\phi
{\cal G}}{\sqrt{1+(\partial_{\rho}\sigma_0)^{2}}}\partial_\rho
g(\rho)\frac{w^4}{\sqrt{(w^4+1)(w^4-1)}}\right)=0\eeq

There is clearly a solution with $M=0$ and $g(\rho)=$constant.
This corresponds to introducing a background gauge field
associated with U(1) baryon number in the field theory. Note that
asymptotically in the UV the lagrangian for this field is

\beq {\cal L} \simeq {N \over 4 \pi^2}
\log{\frac{\Lambda_{UV}}{\Lambda_b}}~ \frac{1}{4} F^{\mu \nu}
F_{\mu \nu} \eeq
 which reflects the
logarithmic running of the flavour gauge coupling.

There is a second asymptotic solution of the equation of motion
where $g(r) \sim 1/\rho^2$ and these solutions correspond to the
vector meson spectrum associated with the operator $\bar{q}
\gamma^\mu q$. By seeking smooth solutions of the equation of
motion we can determine the vector meson mass spectrum. The
results of analysis are shown in the table compared to those
calculated in the pure AdS background. The spectrum (of singlets
on the $S_3$) in the pure AdS case is given by \cite{Mateos}: \beq
M_{v}^2 R^2=4(n+1)(n+2) \eeq

\begin{figure}[!h]
\begin{center}
\begin{tabular}{||l|l|l||}\hline
\emph{n} & \emph{AdS case} & \emph{CM case}\\ \hline 0 & 2.83 &
2.16 \\ \hline 1 & 4.90 & 4.85 \\ \hline 2 & 6.93 & 7.05 \\ \hline
3 & 8.94 & 9.20 \\ \hline 4 & 11.0 & 11.3 \\ \hline 5 & 13.0&
13.5 \\ \hline 6 & 15.0 & 15.6 \\ \hline 7 & 17.0 & 17.7 \\ \hline
8 & 19.0 & 19.9 \\ \hline
\end{tabular}
\caption{Vector meson spectrum comparing CM and pure AdS
backgrounds}\label{VM}
\end{center}
\end{figure}

There are a number of interesting questions one might ask with
regards the vector mesons and weakly gauged chiral symmetries. For
example in models of dynamical symmetry breaking such as
technicolour the value of the parameter $L_{10}$ in (\ref{gl}) is
related to the phenomenological $S$ parameter which is well
measured by the precision electroweak data. This is an UV finite
quantity which in this model would be related to the difference
between the coefficient of the vector $F^2$ term calculated above
and the equivalent term for the axial gauge field. Unfortunately
the DBI action has no field that corresponds to the axial gauge
field. Presumably equally the operators $\bar{q} \gamma^\mu
\gamma_5 q$ are described by a string mode not present in the DBI
action, so it is therefore not possible for us to estimate this
parameter. Similarly it would have been nice to have tested the
Weinberg sum rules \cite{Wein} such as, assuming vector meson
dominance,

\beq f_\Pi^2 = { g_V^2 \over M^2_V} - {g_A^2 \over M_A^2} \eeq
where $g_{V/A}$ are the couplings of the lightest vector and axial
vector mesons to their respective currents. We have no description
of the axial vector mesons though. It would be productive to
understand this sector in the future.

These issues have been highlighted in \cite{Son} where a toy model
of large $N$ QCD was proposed inspired by the general structure of
the AdS/CFT Correspondence. That model has a flavour gauge field
living in a (deconstructed) finite fifth dimension with symmetry
breaking boundary conditions. They interpret $A^5$ as the pion
field and the KK modes of the flavour gauge field as the vector
and axial vector mesons. This is distinct from our scenario where
the pion fields are described by an additional scalar field and
the flavour gauge field only describes vector mesons - the two
pictures do not support each other therefore.

\section{Summary}

We have analyzed chiral symmetry breaking via the gravitational
dual construction of D7 brane probes in the Constable Myers
non-supersymmetric geometry. The ground state of the D7 brane
indeed describes chiral symmetry breaking and a vacuum manifold
with massless pions in the chiral limit. We have shown that these
pion fields are described by a chiral lagrangian in the infra-red
as expected. We have computed the couplings of the low energy
lagrangian and shown that they match expectations from naive
dimensional analysis.

One might have expected significant deviations in the magnitudes
of parameters because the gauge coupling of the model remains
strong in the UV (as in for example a ``walking" gauge theory
\cite{Holdom}). However, the UV of the theory has an approximate
${\cal N}=4$ supersymmetry that forbids a chiral condensate so
that above our symmetry breaking scale $\Lambda_b$ the condensate
decreases rapidly in spite of the strong coupling. The chiral
symmetry breaking behaviour of this model is therefore rather
pleasingly analogous to that of QCD.

\vspace{1cm}

\noindent {\bf \Large Acknowledgements}

We would like to thank Sekhar Chivukula, Rob Myers and David
Matteos for useful discussions.

NE is grateful for the support of a PPARC Advanced Fellowship and
JS for the support of a PPARC studentship.

\end{document}